\documentclass{article}

\PassOptionsToPackage{square,sort,comma,numbers}{natbib}

\usepackage[preprint]{neurips_2021}

\usepackage[utf8]{inputenc} 
\usepackage[T1]{fontenc}    
\usepackage{hyperref}       
\usepackage{url}            
\usepackage{booktabs}       
\usepackage{amsfonts}       
\usepackage{nicefrac}       
\usepackage{microtype}      
\usepackage{xcolor}         

\usepackage{multirow}

\usepackage{caption}

\usepackage{amsmath,amssymb,amsfonts}

\usepackage{graphicx}

\usepackage{subcaption}

\title{Timbre Transfer with Variational Auto Encoding and Cycle-Consistent Adversarial Networks}

\author{%
  Russell Sammut Bonnici \\
  School of Electronic Engineering and Computer Science \\
  Queen Mary University of London\\
  United Kingdom, E14NS \\
  \texttt{ec20074@qmul.ac.uk} \\
   \And
   Charalampos Saitis \\
   School of Electronic Engineering and Computer Science \\
  Queen Mary University of London \\
  United Kingdom, E14NS \\
  \texttt{c.saitis@qmul.ac.uk} \\
   \AND
   Martin Benning \\
   School of Mathematical Sciences \\
   Queen Mary University of London \\
   United Kingdom, E14NS \\
   \texttt{m.benning@qmul.ac.uk} \\
}

\begin{document}

\maketitle

\begin{abstract}
  This research project investigates the application of deep learning to timbre transfer, where the timbre of a source audio can be converted to the timbre of a target audio with minimal loss in quality. The adopted approach combines Variational Autoencoders with Generative Adversarial Networks to construct meaningful representations of the source audio and produce realistic generations of the target audio and is applied to the Flickr 8k Audio dataset for transferring the vocal timbre between speakers and the URMP dataset for transferring the musical timbre between instruments. Furthermore, variations of the adopted approach are trained, and generalised performance is compared using the metrics SSIM (Structural Similarity Index) and FAD (Frech\'et Audio Distance). It was found that a many-to-many approach supersedes a one-to-one approach in terms of reconstructive capabilities, and that the adoption of a basic over a bottleneck residual block design is more suitable for enriching content information about a latent space. It was also found that the decision on whether cyclic loss takes on a variational autoencoder or vanilla autoencoder approach does not have a significant impact on reconstructive and adversarial translation aspects of the model.
\end{abstract}

\section*{Acknowledgments}
The research work disclosed in this publication is funded by the ENDEAVOUR Scholarships Scheme (Malta). The scholarship is part-financed by the European Union – European Social Fund (ESF) under Operational Programme II – Cohesion Policy 2014-2020, “Investing in human capital to create more opportunities and promote the well-being of society”. We would like to express our gratitude to Ehab Albadawy for sharing his source code, feedback, and giving advice that helped resolve a number of conceptual confusions. We would also like to thank Ben Hayes and Cyrus Vahidi from Queen Mary University of London for providing us with insights on works in this problem domain.

\section{Introduction}
Timbre transfer is a task concerned with modifying audio samples such that their timbre is reformed while their semantic content is persisted. Through this, utterances of a speaker (referred to as the source) can be changed such that they sound like they were spoken by another speaker (referred to as the target). Recordings of a source instrument can be manipulated in a similar way such that they sound like another target instrument played them. Applications of effective timbre transfer would benefit areas such as voice anonymisation, music production, and data augmentation. The challenge in making the modification take place first lies in how exactly timbre can be captured.

\par
Timbre is formally defined as the quality of an audio stimulus in which a listener can distinguish two sounds with a factor separate from loudness and pitch \cite{american1976usa}. As reflected by how its definition describes what it is not, timbre is highly abstract and hard to determine concisely. Despite sometimes getting referred to as tone colour, it is harder to quantify than visual colour. Visual colour is commonly defined in three dimensions with an RGB model, and though previous research has determined a three-dimensional model for timbre \cite{mcadams1995perceptual}, it is still not as clear cut.
\par
Explicit characteristics such as spectral envelope and time envelope help determine timbre for instruments, but there are still implicit characteristics that contribute to painting the complete picture. Also, musicians with more exposure to instruments of varying timbre are better at identifying them \cite{saldanha1964timbre}, indicating a direct proportionality between exposure and timbral understanding. These points motivate the use of deep learning, where  from data, hard-to-define patterns of timbre can be learnt by models non-linearly, and such models can be applied for related discriminative and generative tasks.   
\par
Generative modelling is a task that has been increasingly getting more attention in recent years. Like discriminative modelling, intrinsic patterns about a collection of samples are learnt. Unlike discriminative modelling, it does not deduce conclusive information about the samples but uses the learnt patterns to generate new samples for a target sample distribution. From the research field of computer vision, a variety of performant generative models have been proposed for tasks such as data generation and style transfer. Most recent models extend from Variational Autoencoders \cite{kingma2014auto} and Generative Adversarial Networks \cite{goodfellow2014generative}. 
\par
Generative Adversarial Networks (GANs) are a hybrid approach to generative modelling that aim to achieve realistic results with a discriminative overseer. GANs consist of two networks referred to as a generator and discriminator. The generator is concerned with generating sufficiently realistic data such that when it is compared to target data by the discriminator, it is indistinguishable. Meanwhile, Variational Autoencoders (VAEs) take on a reconstructive approach. VAEs are split into two networks referred to as the encoder and decoder. Together, they compose a unified hourglass-like architecture. The encoder learns to compress input data into a highly abstract latent space at a bottleneck central to the model. Since the autoencoder is variational, it learns to model latent data such that it resembles a specified data distribution (typically Gaussian) for a better consistency in description at the bottleneck. For this, a component called Kullback Leibler divergence is used to estimate the log difference between the probability of data in the predicted distribution and the probability of data in the desired distribution. Lastly, the decoder learns to decompress latent data from the bottleneck to the dimensional space of input data. 
\par
In training, VAEs are more stable than GANs. Since VAEs have weights tuned with respect to loss computations of their own output, their optimal state lies within a local minimum. Like most deep learning models, the loss minimises asymptotically. Meanwhile, GANs tune the generative component based on loss computations from a separate adversarial discriminator network. This makes their optimal state lie at a saddle point, introducing more fluctuation to the loss minimisation.


\par
The instability of GANs makes them sensitive to data and design decisions, as well as susceptible to mode collapse. This is a typical failure case where the generator learns to ``cheat'' the discriminator by mapping more than one input sample to the same output. Despite their notorious instability, when configured right, GANs can achieve significantly realistic results. Their vulnerability to mode collapse suggests that they would benefit from architectural decisions in the generator that better capture the latent meaning of input samples. An approach to this would be to remodel the generator into a VAE, yielding a VAE-GAN model design. By combining the two generative techniques, the adversarial aspect becomes more stabilised, while the reconstructive aspect benefits from more motivation for realism.

\vspace{0.34cm}

\section{Related Work}

\subsection{Image Style Transfer}
\par
UNIT \cite{liu2017unsupervised} is a notable model that motivates the adoption of a VAE-GAN approach for style transfer. It focuses on transferring styles of a source image to that of a target image. For a pair of style domains, they train the transfer in both directions. For each transfer direction, they use an encoder-decoder-discriminator pathway. The encoder-decoder section follows the reconstructive objective of VAEs (motivating content persistence). Meanwhile, the whole path with the discriminator included follows the adversarial objective of GANs (motivating style transfer). At the bottleneck, they enforce a shared latent space by computing the VAE objective in both directions. They also share weights of the last layer of the encoders and share the weights of the first layer of the decoders for motivating a shared latent space. 
\par
The reconstructive component of the VAE objective in UNIT was computed using an error criterion L1, which computes the total difference between the absolute magnitudes of a reconstructed image and its original version. Similarly, CycleGAN \cite{zhu2017unpaired} trains in both directions and uses L1 for comparing the quality of a reconstruction to the original input. Both models are cyclically consistent since the cyclic L1 loss acts as a prior that allows the applicability of style transfer to unpaired data, where the content between two images is different. The main difference is that CycleGAN does not assign a VAE objective to the generator section, and so in the cyclic loss, there is no inclusion of Kullback–Leibler divergence alongside the L1 reconstructive component. As a result, CycleGAN learns how to transfer styles from a lower level.
\par
The shared latent space of UNIT aims to address style transfer from a probabilistic modelling perspective. They reason that the goal is to capture the joint distribution of two style domains in order to transfer between them. It is tough to do this when data is unpaired and not captured at a high enough level. The shared latent space aims to better capture the joint distribution by emphasising source content and deemphasising source style. Intuitively, it represents content with less individuality at the bottleneck so that it becomes easier for the decoder to introduce target style to it.



\subsection{Timbre Transfer}

\par
A number of works concerned with audio style transfer burrow intuition from image style transfer models. Timbre transfer is a subset of audio style transfer, for which timbre is taken as the style of interest in the audio domain. Here, there are typically two different design paradigms that researchers follow. They either follow a time-domain approach, where an end-to-end deep learning model deals with audio directly and at a low level. Alternatively, they may follow a time-frequency procedure, where audio is handled more indirectly but at a higher level. In this approach, the data is further processed for less complexity, with two deep learning models used for a high-quality transfer. The first model is concerned with performing style transfer on spectrogram representations of the audio, and the second model is concerned with vocoding the results of generated spectrograms back to realistic audio.
\par
AutoVC \cite{qian2019autovc} deals with utterances of speakers in the time-frequency domain and proposes a style transfer model that follows a vanilla autoencoder architecture. Like UNIT, they motivate content persistence in the bottleneck and style adaptation in the decoder but achieve it without an adversarial component. Their model consists of two encoders; a content encoder and a style encoder. The content encoder focuses on embedding the content of source utterances, whereas the style encoder focuses on embedding the style of target utterances. The decoder then accepts both content and style codes as input so that its transferred output is the amalgamation of source content and target style. The purpose of the style encoder replaces the purpose of the adversarial component in UNIT for target motivation. Furthermore, they focus on mel-spectrogram representations and use WaveNet \cite{oord2016wavenet} to convert generated mel-spectrograms to audio. 
\par
TimbreTron \cite{huang2018timbretron} focuses on recordings of instruments in the time-frequency domain.  They also use WaveNet for vocoding and follow the approach of CycleGAN for their style transfer model. Initially, they perform the style transfer on vanilla spectrograms computed using a Short Time Fourier Transform (STFT) but report issues with correctly transferring low pitches as well as having output pitches generally randomly vary to a degree. To overcome these issues, they apply the style transfer to CQT spectrogram representations instead, where a higher resolution for lower frequencies is captured and pitch equivariance is maintained. Though an improvement, their generated results still lack in quality. Their change of focus to CQT spectrograms likely tries to make up for the lack of latent representation capabilities in CycleGAN. Rather than adopting a different spectrogram representation, it may be better to adopt a VAE-GAN approach such as UNIT that still involves cyclic consistency but can also embed content at a higher level. 

\par
Mor et al \cite{mor2018universal} proposed a model in the time domain for universal music translation.  Their style of interest involves timbre but also extends to a broader aspect, inclusive of composition style. They use a denoising autoencoder architecture, where input data is augmented and learnt to be reconstructed with respect to non-augmented data. The encoding and decoding components are designed following the principles of WaveNet since it excels at generation in the time domain. For the augmentation, pitch is randomly varied within a deviation of 0.5 half-steps. They motivate this so as to capture content representations at a higher level in the bottleneck. Though it yields an improvement in content generalisation, the augmentation may contribute to off-key pitch in the output, which is especially undesirable in musical contexts. With respect to the bottleneck, they apply an efficient shared latent space constraint by using the same encoder for all source domains. Motivating the generalisability of a universal encoder, they achieve applicability to source domains that were unseen from training. They also use a domain confusion loss to discourage the inclusion of style elements in the content embeddings. 
\par
AlBadawy and Lyu \cite{AlBadawy2020} proposed a model for voice conversion in the time-frequency domain. Their style transfer model serves as an extension of UNIT, where a shared latent space is encouraged with a universal encoder (much like Mor et al). They also introduce a latent loss to penalise significantly different averages of latent codes originating from other style domains (also similar to the purpose of the domain confusion loss in Mor et al). It encourages embeddings to be more independent of style by making content distributions align closer in the bottleneck. Their adaptation of Mor et al's intuition to the time-frequency domain eases training expenses as well as architectural complexity. A more semantically rich content embedding may also be achieved with the variational autoencoding elements of UNIT. Much like other related works, they apply their style transfer to mel-spectrograms and use WaveNet for vocoding back into audio. 
\par
This project adopts the VAE-GAN approach taken by AlBadawy and Lyu as their work demonstrates highly realistic results for the voice conversion between two speakers. Investigating potential model improvements as well as extending its applicability to the context of instruments poses an interesting study case. Like speech conversion, timbre remains the main style of interest, but for instruments, the content of interest becomes melodic sequences rather than linguistic sentences. Their time-frequency approach also makes the training and comparing of experiments more feasible both in terms of time and resource requirements. By decoupling the audio generation process (spectrogram vocoder) from the style transfer process, different style transfer experiments may be trained without having to retrain the audio generation process each time. Also, this design is not limited to a specific approach for audio generation, making it easily applicable to a variety of vocoders.

\vspace{0.34cm}

\section{Method}

\subsection{Data Preprocessing}

\par
Prior to computing mel-spectrograms for the style transfer model, the input audio is preprocessed in a number of steps. Firstly, the audio is ensured to match a specified sample rate of 16,000 Hz. If not, then the audio is resampled accordingly. The volume is then subtly equalised with root mean square normalisation, where if the root mean square amplitude of the audio is lower than a target amplitude of -30dB, then the audio is normalised to that amplitude. Lastly, long silences from the audio are masked out to remove background noise at segments where no sound events are present. Note that the low amplitude of -30dB was chosen arbitrarily, subtly equalising low segments of the audio as a result. Higher amplitudes such as -10db may be set in future work. For each of the processed audio samples, mel-spectrograms are then computed with 128 mel frequency bins, a hop size of 200 samples, and a Hanning window of size 800 samples for each frame. Also, magnitude values for each mel-spectrogram are logarithmically scaled and normalised with min-max normalisation for a faster convergence in training. For each audio source (whether speaker or instrument), whole audio and corresponding mel-spectrogram samples are organised into subsets of 80\% for training, 10\% for validation, and 10\% for testing.


\subsection{Components and Data Input}

\begin{figure}[h!]
    \centering
    \includegraphics[width=\columnwidth, height= 2.7cm]{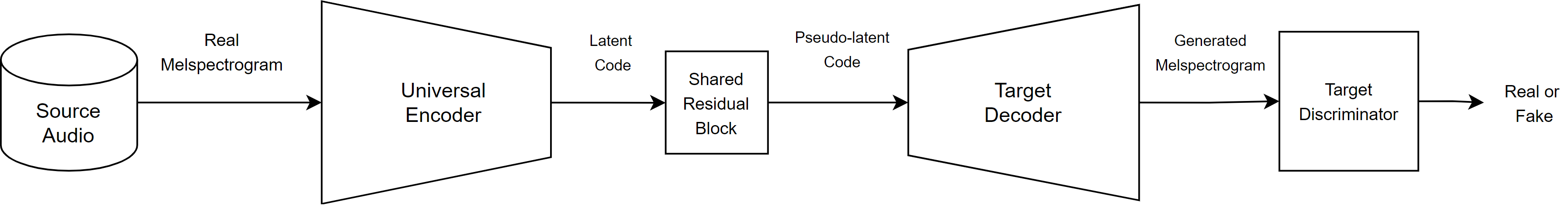}
    \caption[width=\textwidth]{The design of a single path in the implemented VAE-GAN model, where the transfer of one timbre to another is learnt in a mel-spectrogram format. In training, two paths allow a one-to-one style transfer between timbres in both directions. For each transfer direction, the encoder aims to extract the content of the source audio regardless of timbre, whereas the decoder aims to introduce the timbre of the target audio to the content as motivated by the discriminator.}
    \label{fig:one-path}
\end{figure}

\par
The style transfer model involves one encoder, a shared residual block for decoding at a shallow level, and multiple decoder-discriminator pairs specific to target domains. An illustration of a single path within this model is presented in Figure \ref{fig:one-path}. Only two paths are utilised for a one-to-one style transfer in the initial implementation, where for a forward path that converts timbre A to timbre B, there also exists an inverse path that converts B to A. This is necessary for cyclic consistency. For each source, only one decoder-discriminator pair is trained for a target. This design may be extended to many-to-many style transfer by having each source simultaneously train different decoder-discriminator pairs for multiple targets. Not only would this make it easier to extend the model to numerous examples of timbre transfer, but the shared encoder would further benefit in terms of generalisation.

\par
The input mel-spectrograms are subsamples of the full mel-spectrograms such that their width corresponds to a length of 128 frames (1.6 seconds), making the input mel-spectrograms of a resolution size of 128x128. This was necessary to have a standardised input size since the model largely depends on convolutional operations. For each of the domains, a combination of samples was randomly selected from the training set. From each sample, a 128 frame excerpt is extracted between random frame indices. This achieves a random data selection across domains and in terms of time localisation within each mel-spectrogram. 

\subsection{Architecture} 

\par
The architectural details of the model's components are primarily congruent to the design choices initially proposed by UNIT \cite{liu2017unsupervised}. The shared encoder involves two principal stages; a convolutional stage and a residual stage. It firstly pads input with a 3x3 reflection padding. After which, padded information is passed through 3 convolutional blocks that each downsample feature maps by a factor of 2. The first input channel is set to 1 to correspond to the loudness dimension of the mel-spectrogram. The 3 convolutional blocks then have output channels set at 32, 64, and 128 consecutively such that the feature map dimensions are stretched in-depth and compressed in width and height (leading to a compressed latent space). Each convolutional layer is followed by a LeakyReLu activation with an 0.2 negative slope and followed by an instance normalisation layer. The first convolutional layer has a kernel size of 7x7 whereas the following two have 4x4 kernel sizes. Following the convolutional blocks are 3 residual blocks. 
\par
The residual blocks involve skip connections that enrich the representation of data leading to the bottleneck with residual priors. The skip connections are implemented such that for each block, the input feature map is connected to the output feature map via a summation operation. The initially implemented blocks correspond to the basic residual block design proposed by ResNet \cite{he2016deep}, where each block consists of two convolutional layers of a 3x3 kernel size and only the first layer has ReLu applied after it. Unlike the ResNet design, here input is padded with a reflection padding of 1x1 and instance normalisation is applied after the convolutional layers instead of batch normalisation.
\par
Instead of using basic residual blocks, bottleneck residual blocks (also proposed by ResNet) may be considered. Bottleneck residual blocks contain 3 convolutional layers of kernel sizes 1x1, 3x3, and 1x1. Due to the compression that their 3x3 convolutional layer provides, feature maps are reduced in size and so the operations required by a forward pass are reduced. This would benefit the model in terms of easing training expenses, but it may also make training more susceptible to vanishing gradients since the feature maps around the main bottleneck are already at a sufficiently compressed depth. 
\par
The output of the last encoder residual block is taken as a latent mean $\mu$. The reparameterisation trick is then ensued where a random sample is selected from a zero-mean Gaussian distribution with a unit variance of 1. The sum of the random sample and the latent mean is computed to achieve a latent code $z$, which aims to represent the content of the input mel-spectrogram at the bottleneck.


\par
The latent code $z$ is taken as input to the decoder. The decoder follows the exact same architecture as the universal encoder but in a mirrored manner so that a 128x128 mel-spectrogram can be reconstructed as output. The 3 residual blocks follow the same setup as in the encoder, without much need for a reordering since each block has its output dimensions correspond to their input dimensions. The only difference is that instead of all residual blocks being shared across different source domains, only the first one is. The other two are specific to the target domain. This is meant to help motivate content abstraction, where content is decoded once in a similar manner before getting decoded further with respect to the target. The following 3 convolution blocks are also specific to the target and utilise transposed convolutional layers that consecutively follow kernel sizes 4x4, 4x4 and 7x7 with output channels 64, 32, and 1, respectively.
\par
The discriminator accepts generated mel-spectrograms from the corresponding target decoder and compares them with real mel-spectrograms of the target domain. Its design is similar to the discriminator proposed by DCGAN \cite{dcgan}. It consists of five convolutional layers, each with a stride of 2x2 so that pooling is performed implicitly and optimally. The first four convolutional layers have a kernel size of 4x4 and the last has a kernel size of 3x3. All layers except the last are followed by a LeakyRelu of 0.2, and the second to fourth layers are each followed by instance normalisation.


\subsection{Overall Objective and Loss Functions}
The overall loss function $L$ that we aim to minimise is additively composed of four individual loss functions, i.e.

\begin{equation}
\label{objective}
    L =  
    L_{\text{GAN}} +
    L_{\text{VAE}} + 
    L_{\text{CC}} + 
    L_{\text{Latent}} \, ,
\end{equation}

\par
where the individual loss functions $L_{\text{GAN}}$, $L_{\text{VAE}}$, $L_{\text{CC}}$ and $L_{\text{Latent}}$ will be defined throughout the remainder of this section. In training, the weights of the generator sub-paths (encoder-decoder) and the weights of the discriminators are updated simultaneously. The generators are tuned with respect to  reconstructive losses for content embedding ($L_{\text{VAE}}$, $L_{\text{CC}}$ and $L_{\text{Latent}}$), and an adversarial loss for trying to generate realistic mel-spectrograms in the target domain. Meanwhile, discriminators depend on an adversarial loss but from the perspective that opposes the generators, where weights are tuned to distinguish generated mel-spectrograms from real mel-spectrograms in the target domain. From these objectives, an overall objective is implied as described by Equation \ref{objective}.

The adversarial loss $L_{\text{GAN}}$ is defined as
\begin{equation}
\label{generator}
    L_{\text{GAN}} = \lambda_0 \mathbb{E}_{x\sim X} (\log D(x)) + \lambda_0 \mathbb{E}_{z\sim Z}(\log(1 - D(G(z)))) \, .
\end{equation}


 In this notation, $D(x)$ refers to the classification output of the discriminator. For each discriminator, an error is minimised such that a mel-spectrogram x (whether fake or real) is correctly identified as belonging to their prior distribution $X$. Meanwhile, for each generator, an error is minimised such that a mel-spectrogram $G(z)$ generated via a latent code $z$ (with a prior of the latent distribution $Z$) is identified as real. The discriminators are assessed on classifying mel-spectrograms of the target timbre to enforce realistic target motivation.
\par
As opposed to the initial definition proposed by Goodfellow et al \cite{goodfellow2014generative}, $L_{\text{GAN}}$ does not use Binary Cross-Entropy (BCE) for the error criterion but instead uses Mean Squared Error (MSE) as motivated by LSGAN \cite{mao2017least}. Between the predicted classifications and ground truth labels, BCE computes the logarithmic difference of probabilities, whereas MSE computes the total difference of squared magnitudes. MSE is typically used in recent style transfer models as it calculates the difference between classifications and desired classifications less discretely and further stabilises the adversarial aspect of the training process. From adopting MSE over BCE, it is worth noting that no sigmoid activation is required as a final layer in the discriminator. 


The variational encoding loss $L_{\text{VAE}}$, i.e.
\begin{equation}
\label{VAE}
    L_{\text{VAE}} = \lambda_1 \mathbb{KL}(Z, p(z)) - \lambda_ 2 \mathbb{E}_{z\sim Z}(\log p_D(x|z))
\end{equation}
consists of two terms. The first term is Kullback Leibler (KL) divergence between the latent distribution $Z$ and a zero-mean Gaussian distribution $p(z)$. The latent distribution $Z$ is defined by the output $\mu$ of the encoder. The second term is the reconstructive component that aims to successfully recover a mel-spectrogram $x$ from a corresponding latent code $z$ through a probabilistic decoder $p_D(x|z)$. For this, L1 is used as an error criterion as it encourages sparsity which is suitable for mel-spectrograms. L1 is computed between the input source mel-spectrogram and the reconstructed source mel-spectrogram (recovered by feeding $z$ to a decoder from an inverse path).

The cyclic consistency loss $L_{\text{CC}}$, i.e.
\begin{equation}
\label{CC}
    L_{\text{CC}} = \lambda_3 \mathbb{KL}(Z_{cc}, p(z_{cc})) - \lambda_4 \mathbb{E}_{z_{cc}\sim Z_{cc}}(\log p_D(x|z_{cc})) \, ,
\end{equation}
 is computed with the same loss components as $L_{\text{VAE}}$ but the estimated latent distribution $Z_{CC}$ and latent code $z_{cc}$ are taken from a cyclic reconstruction. Here, a generated mel-spectrogram is encoded again to obtain $Z_{CC}$ and $z_{cc}$, and by using the decoder from an inverse path, the source mel-spectrogram is reconstructed. The right term ensures cyclic consistency. Meanwhile, the left term makes it so that the latent space distribution gets modelled with respect to generated mel-spectrograms. Since the encoder is shared and the same weights are tuned for both $Z$ and $Z_{CC}$, this may serve as an obstacle in modelling a latent space distribution at the bottleneck specific to real mel-spectrograms. It may be worth investigating a model variation where the KL divergence is omitted in $L_{CC}$.

\begin{equation}
\label{Latent}
    L_{\text{Latent}} = \lambda_5 | \mu_A - \mu_B |
\end{equation} 

The latent loss $L_{\text{Latent}}$ (Equation \ref{Latent}) is the L1 error between a pair latent means $\mu_A$ and $\mu_B$ from different source mel-spectrograms A and B respectively. This makes it so that a focus on embedding content is further encouraged at the bottleneck. In the case that a many-to-many variation of this model is pursued, $L_{\text{Latent}}$ would have to be calculated between each possible pairing of latent means.
\par
The constant hyperparameters for the loss functions were set such that $\lambda_0=10, \lambda_1=0.1, \lambda_2=100, \lambda_3=0.1, \lambda_4=100, \lambda_5=10$. Here, the reconstructive loss components are favoured over the rest to provide more stability in loss minimisation and to prioritise encoding content before focusing on decoding target mel-spectrograms. Moreover, Adam optimisers \cite{kingma2014adam} were used with a set learning rate of $\alpha=0.0001$ and running average coefficients $\beta_0=0.5$ and $\beta_1=0.999$. Learning rate schedules were used to start decaying the learning rate from halfway through the maximum epochs for a given dataset. For training, the batch size was set to 4 samples per timbre domain.
 
\subsection{Inference}

After the model is trained, an inference procedure is set up such that it can be applied to input audio of an arbitrary length (provided a length longer than 1.6 seconds). The audio is first preprocessed and a mel-spectrogram is computed with the same method used for training. Since the model only accepts 128x128 mel-spectrograms, a sliding window of a 128 frame length is set to traverse the audio with an overlap count of 4. With each slide, a mel-spectrogram is inferred with the timbre of the target audio. An average of the magnitude values is then taken between the overlapping regions of the transferred mel-spectrograms, resulting in the timbre transfer of a full-length sample. 
\par
After the inference, the Fast Griffin Lim algorithm \cite{perraudin2013griffinFast} is used to convert the inferred mel-spectrogram to an audio format. Since Griffin Lim is susceptible to phase artefacts, the reconstructed audio will be taken as input to the mel-spectrogram-to-audio vocoder model (in this case, WaveNet) for quality improvement. Here, the reconstructed audio would be preprocessed back to mel-spectrograms with respect to the foreign vocoder's preprocessing steps, then reconstructed once more back to audio but with much fewer phase artefacts. This makes it so that it is not a requirement for the preprocessing specifications of the style transfer model to match that of the vocoder.

\par

\subsection{WaveNet Vocoding}
\par
An open-source implementation for WaveNet was utilised \cite{r9y9} and separately trained for each of the datasets. One was trained for converting timbres between speakers, and another for converting between instruments. Each vocoder was trained on all timbres per dataset since intonations that may only be present in the source domain and not the target domain should still be considered for style transfer. With dilated convolutions and causal filters, the vocoders successfully learn to generate audio from preprocessed mel-spectrograms and minimise the loss in the audio reconstruction. This results in models of different weights for audio generation general to the timbres from each task domain, which are then used for inferring high-quality audio reconstructions of the target audio generations from the style transfer model.

\vspace{0.34cm}

\section{Experiments and Evaluation}

\subsection{Datasets}

The model was trained separately on two datasets; the Flickr 8k Audio dataset \cite{flickr}, and the URMP dataset \cite{urmp}. From the Flickr dataset, audio files of male and female speakers with the most recordings were used for investigating model variations in the same context as AlBadawy \& Lyu \cite{AlBadawy2020}. Here, speakers utter a variety of short sentences. Meanwhile, from the URMP dataset, audio files of instruments with the most to least recordings were used for extending the model's application to the context of musical timbre. To mirror the application of voice conversion, only solo recordings of instruments were considered where timbre is monophonic and not polyphonic, though experiments with polyphonic timbre (like in Mor et al \cite{mor2018universal}) may be worth investigating in the future.

\begin{table}[h!]
\centering
\caption{Dataset information with respect to each timbre of interest}
\label{tab:timbre-count}
\begin{tabular}{|c|l|c|c|c|}
\hline
\textbf{Dataset} &
  \multicolumn{1}{c|}{\textbf{Source}} &
  \textbf{Samples} &
  \textbf{Total Time} &
  \textbf{\begin{tabular}[c]{@{}c@{}}Avg. Time \\ (per sample)\end{tabular}} \\ \hline
\multirow{4}{*}{\begin{tabular}[c]{@{}c@{}}Flickr\end{tabular}} &
  Female 1 &
  1686 &
  1 hrs, 48 mins, 1 s &
  3.6 s \\ \cline{2-5} 
                      & Male 1   & 2965 & 2 hrs, 46 mins, 36 s & 3.6 s         \\ \cline{2-5} 
                      & Female 2 & 1058    & 58 mins, 9 s                    & 3.6 s             \\ \cline{2-5} 
                      & Male 2   & 2461    & 2 hrs, 30 mins, 43 s                    & 3.6 s             \\ \hline
\multirow{4}{*}{URMP} & Trumpet  & 22   & 35 mins, 36 s        & 1 min, 32.4 s \\ \cline{2-5} 
                      & Violin   & 34   & 51 mins, 1 s         & 1 min, 20.4 s \\ \cline{2-5} 
                      & Flute    & 18    & 28 mins, 30 s                    & 1 min, 50 s             \\ \cline{2-5} 
                      & Cello    & 11    & 16 mins, 50 s                    & 1 min, 57 s             \\ \hline
\end{tabular}
\end{table}

\par
The number of samples per timbre and time length information were summarised in Table \ref{tab:timbre-count}. Speakers from Flickr have much more samples than instruments from URMP. On the other hand, the average time per sample is much shorter for Flickr than URMP. Still, the total time recorded of URMP instruments still amounts to significantly less than Flickr speakers. As a result, the URMP experiments should be trained for more epochs such that the total number of steps better match the amount computed for Flickr. Furthermore in Flickr, male 1 has the largest amount of samples whereas female 2 has the smallest. Meanwhile in URMP, violin has the most samples, whereas cello has the least.

\subsection{Metrics}

\par
Two metrics were utilised for evaluating the two main aspects of the model; SSIM (Structural Similarity Index \cite{ssim}) for the reconstructive aspect, and FAD (Frech\'et Audio Distance \cite{kilgour2018fr}) for the adversarial translation aspect. Here, SSIM focuses on judging reconstruction in terms of mel-spectrograms of the style transfer model and FAD focuses on comparing generated target audio (after WaveNet) with real target audio.
\par
SSIM is a similarity metric that compares the perceptual quality between an original image and its reconstructed counterpart. This is typically used for assessing image compression algorithms. This is applicable here since mel-spectrograms are two-dimensional and VAEs depend on reconstructing samples post-compression. SSIM is more suitable over other metrics such as PSNR (Peak Signal Noise Ratio) or MSE as it better considers structure \cite{ssim-compare} which is especially important in time-frequency representations. By using SSIM to compare a reconstructed mel-spectrogram with its original (after one encoding pass) and separately a cyclic reconstruction with its original (after two encoding passes), the two reconstructive processes of the model are assessed. 
\par
FAD takes the approach of FID (Frech\'et Inception Distance \cite{fid}) and adapts it from the context of images to audio. It uses a VGGish model \cite{vggish} (a variant of the discriminative model VGG \cite{vgg}) that is pre-trained on a large-scale audio event dataset called AudioSet \cite{audioset}. Due to the dataset it is trained on, its weights are able to produce semantically rich embeddings of audio. FAD estimates the multivariate Gaussians of VGGish embeddings of real audio samples and separately generated audio samples, then computes the Frech\'et Distance between them. This effectively estimates the difference between the two data distributions where the smaller the distance, the more realistic the generated set of samples are. As a computed metric, FAD is found more favourable over person dependent metrics such as MOS (Mean Opinion Score) since it is more objective and better replicable.

\subsection{Model Experiments}

\par
Four versions of the style transfer model were trained for evaluation. The initial version follows the proposed specifications of AlBadawy and Lyu \cite{AlBadawy2020}. Meanwhile, the no KLD cyclic version makes it so that KL divergence is not included in the cyclic loss component, making the model focus more on real input for modelling the distribution of the shared latent space. The bottleneck residual version replaces all basic residual blocks with bottleneck residual blocks to investigate the effectiveness of an alternate design with fewer parameters. And finally, the many-to-many version introduces more pathways in training for cyclically transferring between more timbre domains (in this case four) which should further enforce and generalise a shared latent space.
\par
For the Flickr dataset, most style transfer models were trained for 100 epochs and for the URMP dataset, they were trained for 500 epochs (due to the difference in dataset size). Epochs for training the many-to-many experiments were reduced to 17 for Flickr and 84 for URMP since 6 times the amount of steps were pursued per epoch by each network. For each dataset, a WaveNet vocoder was trained for 450,000 steps since plots seemed to feasibly align with input audio plots at this point (as demonstrated in the Appendix Section C). Most of the experiments were one-to-one where the source domain was taken as the timbre opposing the target. Due to time limitations for training, one-to-one experiments were only pursued for the first two selected timbres of each dataset (between female 1 and male 1 for speakers, and trumpet and violin for instruments). For the many-to-many experiments, metrics were calculated in pairs between the first two selected timbres and separately between the last two selected timbres. Unlike in training, timbres were not exhaustively pursued for the evaluation due to the expensive time requirements posed by the inference procedure of WaveNet. Either with more time or another vocoder with a less time costly inference procedure, more timbre pairings may be investigated for further evaluation. 






\begin{table}[h!]
\centering
\caption{SSIM of Reconstructions}
\label{tab:recon}
\begin{tabular}{|c|l|c|c|c|l|}
\hline
\multirow{2}{*}{\textbf{Dataset}} &
  \multicolumn{1}{c|}{\multirow{2}{*}{\textbf{Target}}} &
  \multicolumn{4}{c|}{\textbf{Model Experiments}} \\ \cline{3-6} 
 &
  \multicolumn{1}{c|}{} &
  \textbf{Initial} &
  \textbf{\begin{tabular}[c]{@{}c@{}}No KLD \\ Cyclic\end{tabular}} &
  \textbf{\begin{tabular}[c]{@{}c@{}}Bottleneck\\ Residual\end{tabular}} &
  \multicolumn{1}{c|}{\textbf{\begin{tabular}[c]{@{}c@{}}Many to \\ Many\end{tabular}}} \\ \hline
\multirow{4}{*}{\begin{tabular}[c]{@{}c@{}}Flickr\end{tabular}}          & Female 1 & \textbf{0.87} & \textbf{0.87} & 0.79 & 0.86 \\ \cline{2-6} 
                                                                         & Male 1   & 0.89 & 0.88 & 0.76 & \textbf{0.91} \\ \cline{2-6} 
                                                                         & Female 2 & - & - & - & 0.89 \\ \cline{2-6} 
                                                                         & Male 2   & - & - & - & 0.87 \\ \hline
\multirow{4}{*}{URMP}                                                    & Trumpet  & 0.93 & 0.93 & 0.85 & \textbf{0.94} \\ \cline{2-6} 
                                                                         & Violin   & 0.91 & 0.91 & 0.82 & \textbf{0.92} \\ \cline{2-6} 
                                                                         & Flute    & - & - & - & 0.90 \\ \cline{2-6} 
                                                                         & Cello    & - & - & - & 0.91 \\ \hline
\end{tabular}
\end{table}

\begin{table}[h!]
\centering
\caption{SSIM of Cyclic Reconstructions}
\label{tab:cyclic}
\begin{tabular}{|c|l|c|c|c|l|}
\hline
\multirow{2}{*}{\textbf{Dataset}} &
  \multicolumn{1}{c|}{\multirow{2}{*}{\textbf{Target}}} &
  \multicolumn{4}{c|}{\textbf{Model Experiments}} \\ \cline{3-6} 
 &
  \multicolumn{1}{c|}{} &
  \textbf{Initial} &
  \textbf{\begin{tabular}[c]{@{}c@{}}No KLD \\ Cyclic\end{tabular}} &
  \textbf{\begin{tabular}[c]{@{}c@{}}Bottleneck\\ Residual\end{tabular}} &
  \multicolumn{1}{c|}{\textbf{\begin{tabular}[c]{@{}c@{}}Many to \\ Many\end{tabular}}} \\ \hline
\multirow{4}{*}{\begin{tabular}[c]{@{}c@{}}Flickr\end{tabular}}          & Female 1 & 0.73 & 0.74 & 0.73 & \textbf{0.77} \\ \cline{2-6} 
                                                                         & Male 1   & 0.80 & 0.78 & 0.68 & \textbf{0.82} \\ \cline{2-6} 
                                                                         & Female 2 & - & - & - & 0.78 \\ \cline{2-6} 
                                                                         & Male 2   & - & - & - & 0.77 \\ \hline
\multirow{4}{*}{URMP}                                                    & Trumpet  & 0.83 & 0.83 & 0.78 & \textbf{0.89} \\ \cline{2-6} 
                                                                         & Violin   & 0.81 & 0.81 & 0.78 & \textbf{0.88} \\ \cline{2-6} 
                                                                         & Flute    & - & - & - & 0.82 \\ \cline{2-6} 
                                                                         & Cello    & - & - & - & 0.85 \\ \hline
\end{tabular}
\end{table}

\par
As shown across a variety of timbres, the reconstruction quality of the first reconstruction (Table \ref{tab:recon}) is consistently higher than the cyclic reconstruction (Table \ref{tab:cyclic}). This indicates a subtle loss of information with each transfer since the cyclic reconstruction goes through one more encoding pass and decoding pass than the first reconstruction. As shown in both tables, most of the investigated VAE-GAN variations do not supersede the SSIM of the initial version, but a majority of the many-to-many experiments do. This validates the hypothesis that content-encoding benefits from having a larger variety of source domains and indicates that with more timbres considered, less information is lost in encoding content. 
\par
The bottleneck residual experiments demonstrate a considerable lack in reconstruction quality, and so basic residual blocks are more suitable for enriching content information to and from the latent space. The no KLD cyclic experiments achieve reconstruction performance on par with the initial experiments, showing that its inclusion or exclusion does not hold a significant impact on the results. This may be due to the adversarial aspect sufficiently motivating translated data to resemble real data such that no obstructions are made when modelling the bottleneck data distribution for the real data.
\par
Inspecting the many-to-many experiments with respect to each dataset, male 1 achieved the best SSIM for both types of reconstructions. This could likely be attributed to male 1 having the largest data size out of all other timbres (Table \ref{tab:timbre-count}). Meanwhile for URMP, trumpet achieved the best SSIM for both types of reconstructions. Across the other one-to-one experiments, it is also found that trumpet still achieves the highest SSIM (if not the equivalent). The fact that trumpet has a higher SSIM than violin is surprising since violin has a larger data size than the other instrument timbres. This may indicate that trumpet has a less complex timbre than violin. 


\begin{table}[h!]
\centering
\caption{Frech\'et Audio Distance (General Vocoding)}
\label{tab:fad}
\begin{tabular}{|c|l|c|c|c|l|}
\hline
\multirow{2}{*}{\textbf{Dataset}} &
  \multicolumn{1}{c|}{\multirow{2}{*}{\textbf{Target}}} &
  \multicolumn{4}{c|}{\textbf{Model Experiments}} \\ \cline{3-6} 
 &
  \multicolumn{1}{c|}{} &
  \textbf{Initial} &
  \textbf{\begin{tabular}[c]{@{}c@{}}No KLD \\ Cyclic\end{tabular}} &
  \textbf{\begin{tabular}[c]{@{}c@{}}Bottleneck\\ Residual\end{tabular}} &
  \multicolumn{1}{c|}{\textbf{\begin{tabular}[c]{@{}c@{}}Many to \\ Many\end{tabular}}} \\ \hline
\multirow{4}{*}{\begin{tabular}[c]{@{}c@{}}Flickr\end{tabular}}          & Female 1 & 2.96 & \textbf{2.77} & 9.10 & 4.31 \\ \cline{2-6} 
                                                                         & Male 1   & 1.65 & 2.48 & 6.97 & \textbf{1.40} \\ \cline{2-6} 
                                                                         & Female 2 & - & - & - & 2.64 \\ \cline{2-6} 
                                                                         & Male 2   & - & - & - & 2.90 \\ \hline
\multirow{4}{*}{URMP}                                                    & Trumpet  & \textbf{5.26} & 5.52 & 6.06 & 5.85 \\ \cline{2-6} 
                                                                         & Violin   & \textbf{4.50} & 5.52 & 12.68 & 4.99 \\ \cline{2-6} 
                                                                         & Flute    & - & - & - & 5.64 \\ \cline{2-6} 
                                                                         & Cello    & - & - & - & 16.21 \\ \hline
\end{tabular}
\end{table}


\par
FADs computed between the generated audio and real audio of each timbre are presented in Table \ref{tab:fad}. For each computation, the target timbre's training set was taken as real audio, whereas the transferred audio (post-processed with WaveNet) was taken as the generated audio to test. The implications of the FAD results for which experiment performed best overall are less clear than the SSIM results, probably due to the introduced dependence on WaveNet. For female 1, no KLD cyclic performed best, while for male 1, many-to-many performed best. On the other hand for the instruments, the initial experiments performed best for trumpet and violin. 
\par
In comparison to the SSIM results, the FAD results between initial and no kld cyclic experiments also do not differ significantly. The bottleneck residual experiments achieved the worst performance for FAD like SSIM. On the other hand, the many-to-many experiments generally did not yield the best performance for FAD like SSIM. It is unclear as to why exactly, but this may be attributed to WaveNet and its potential requirements for training with a more balanced dataset since for the many-to-many experiments, male 1 achieved the best FAD for speakers and violin achieved the best FAD for instruments. This aligns with the fact that male 1 and violin have the most data for Flickr and URMP, respectively (Table \ref{tab:timbre-count}).
\par
Generally, FAD values are worse for instruments than they are for speakers. Also after training the WaveNet vocoders, intonations from mel-spectrograms were reconstructed to less of an accurate degree for instruments (as showcased in Appendix Section C). This could be largely attributed to the lack of instrument data relative to the speaker data (Table \ref{tab:timbre-count}). Further backing this up, Griffin Lim reconstructions of cello audio were surprisingly found to sound more audibly realistic than WaveNet reconstructions. Reflecting the lack of quality, the FAD for cello is significantly worse than the FADs of other timbres. The fact that cello was the timbre with the least data and that no other timbres sounded as bad post-WaveNet implies that WaveNet is sensitive to data size and would greatly benefit from more data for training. Furthermore, audio signals of instruments can be more complex than speaker signals, and so mel-spectrograms may not capture information as well as possible for instruments. Even though results are audibly more realistic than TimbreTron \cite{huang2018timbretron}, it may be worth investigating a VAE-GAN design for CQT spectrograms at least for subtle FAD improvements. Another possible future direction would be to investigate other mel-spectrogram vocoders since by design, WaveNet may be better suited for vocoding the audio of speakers than instruments.

\par

\subsection{Target Visualisations}

\begin{figure}[ht]
    \centering
    \includegraphics[width=\textwidth]{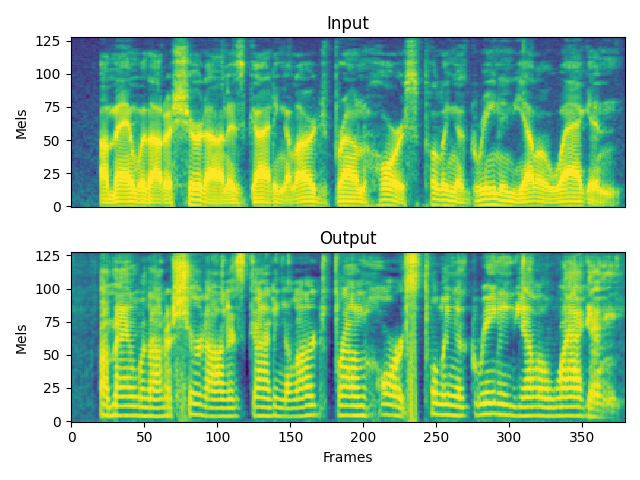}
    \caption{A target visualisation from the many-to-many experiment of female 1, with male 1 as the source input.}
    \label{fig:female1}
\end{figure}

\begin{figure}[ht]
    \centering
    \includegraphics[width=\textwidth]{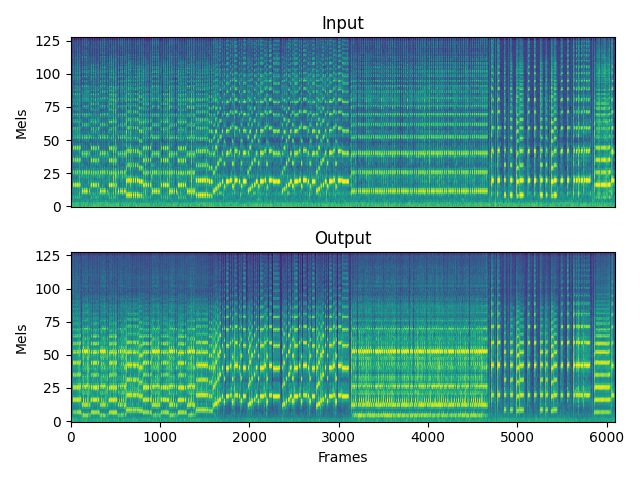}
    \caption{A target visualisation from the many-to-many experiment of trumpet, with violin as the source input.}
    \label{fig:trumpet}
\end{figure}

\par
Input and output mel-spectrograms of targets female 1 and trumpet were plotted to demonstrate the timbre transfer capabilities of the VAE-GAN model as seen in Figures \ref{fig:female1} and \ref{fig:trumpet} (visualisations of other targets may be found in the Appendix Section B). The model retains the content as demonstrated by the vertical spectral features yet modifies the horizontal formants such that they are better suited to the target timbre. From male 1 to female 1, formants are modified such that they are more spaced out and exist less sparsely in a higher range of frequency. From violin to trumpet, formants are introduced at low frequencies, more sparsity is introduced at higher frequencies, and the spacing of formats is modified in a particular way (as can be seen between frames 3,000 and 5,000). These modifications are specific to the target timbres such that their real spectral structure is replicated (of which descriptions and visualisations can be found in Appendix Section A).




\vspace{0.34cm}

\section{Conclusion and Future Work}

\par
In conclusion, a VAE-GAN approach to timbre transfer was not only shown viable in the context of speakers (for voice conversion \cite{AlBadawy2020}{}) but also musical instruments. The instrument timbre transfer results achieved a sufficient audible quality with a relatively simple model working in the time-frequency domain. This model may also be applicable to the transfer of polyphonic timbre in the future since it does not depend on a monophonic pitch transcriptions such as works like \cite{michelashvili2020hierarchical}. The lack of a dependence on a monophonic pitch transcription likely hurts the quality for instrument timbre transfer (as the audible quality of \cite{michelashvili2020hierarchical} is evidently higher), but at least this allows the approach to be general enough for application to more than just one type of audio style transfer problem. With more data as well as further design considerations, the audio quality of this approach may be improved. 
\par
From the VAE-GAN model experiments, a number of indications were deducted across speakers and instruments; that basic residual blocks supersede bottleneck residual blocks around the bottleneck for enriching content information, that the presence of KL divergence for the cyclic loss component does not significantly impact performance, and finally, that the many-to-many extension outperforms the initial one-to-one version in terms of reconstructive capabilities due to the increased variation of data passed through the universal encoder. Though many-to-many improves the reconstructive aspect of the model, improvements on the adversarial translation aspect were inconclusive. More clarity may be produced by training WaveNet with a more balanced dataset and with more data, or by adopting a different time-frequency vocoder with less sensitivity to data quantity. 

\bibliographystyle{unsrt}
\bibliography{main.bib}

\appendix

\section{Appendix}

\subsection{Dataset Visualisations}
\par
\begin{figure}[h!]
\centering
  \begin{subfigure}[b]{0.45\columnwidth}
    \includegraphics[width=\linewidth]{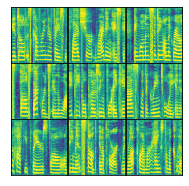}
    \vspace{-18pt}
    \caption{Female 1}
  \end{subfigure}
  \begin{subfigure}[b]{0.45\columnwidth}
    \includegraphics[width=\linewidth]{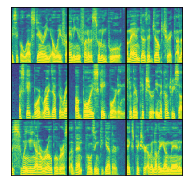}
    \vspace{-18pt}
    \caption{Male 1}
  \end{subfigure} \\
   \begin{subfigure}[b]{0.45\columnwidth}
    \includegraphics[width=\linewidth]{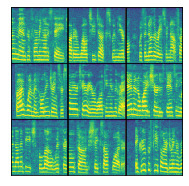}
    \vspace{-18pt}
    \caption{Female 2}
  \end{subfigure}
  \begin{subfigure}[b]{0.45\columnwidth}
    \includegraphics[width=\linewidth]{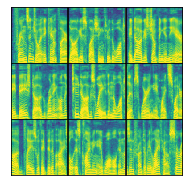}
    \vspace{-18pt}
    \caption{Male 2}
  \end{subfigure}
  \caption{A variety of real mel-spectrogram excerpts per speaker from the Flickr dataset. }
  \label{flickr}
\end{figure}
\begin{figure}[h!]
\centering
  \begin{subfigure}[b]{0.45\columnwidth}
    \includegraphics[width=\linewidth]{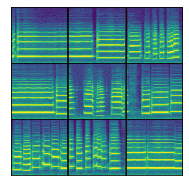}
    \vspace{-18pt}
    \caption{Trumpet}
  \end{subfigure}
  \begin{subfigure}[b]{0.45\columnwidth}
    \includegraphics[width=\linewidth]{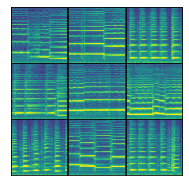}
    \vspace{-18pt}
    \caption{Violin}
  \end{subfigure} \\
   \begin{subfigure}[b]{0.45\columnwidth}
    \includegraphics[width=\linewidth]{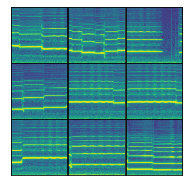}
    \vspace{-18pt}
    \caption{Flute}
  \end{subfigure}
  \begin{subfigure}[b]{0.45\columnwidth}
    \includegraphics[width=\linewidth]{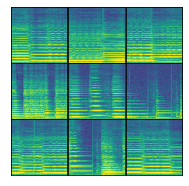}
    \vspace{-18pt}
    \caption{Cello}
  \end{subfigure}
  \caption{A variety of real mel-spectrogram excerpts per instrument from the URMP dataset.}
  \label{urmp}
\end{figure}

A variety of 128x128 mel-spectrogram excerpts were plotted from the selected speakers and instruments to illustrate differences in timbre from the time-frequency perspective of the model (as shown in Figures \ref{flickr} and \ref{urmp} respectively). 
As presented in Figure \ref{flickr}, the horizontal features of females are more vertically spaced out and less compressed to lower regions than that of males. This is reflective of how female voices are higher than males, where the frequency bands exist more prominently over a wider and higher frequency range. There are not many significantly noticeable differences within sex, but female 2 seems to have less intensity in their utterances than female 1. 


\par
In comparison to utterances (Figure \ref{flickr}), the horizontal features of the solo musical performances (Figure \ref{urmp}) are much longer due to the slower pace of the audio content. The horizontal features of violin are more prominent in the upper region than that of the other instruments, indicating a higher voicing. The trumpet has an evident sparseness in the upper area, and so does the flute but to slightly less of a degree. The cello has a dense organisation of horizontal features compressed to the lower region. Similar to how features of males were from Flickr, this is indicative of a lower voicing. At a closer inspection, it is also noticeable that the violin and cello sometimes have subtly oscillating frequency bands. This is indicative of vibrato, a technique commonly used in string instruments.

\subsection{Extra Target Visualisations}

\par
Examples of inferred targets not shown in the main text are presented here in Figures 6-11. By comparing the generated target output to the corresponding mel-spectrograms of real data in Figures \ref{flickr}-\ref{urmp}, it can be seen that the VAE-GAN modifies the nature of the spectral bands such that they better match the specified target.

\begin{figure}[h!]
    \centering
    \includegraphics[width=0.915\columnwidth]{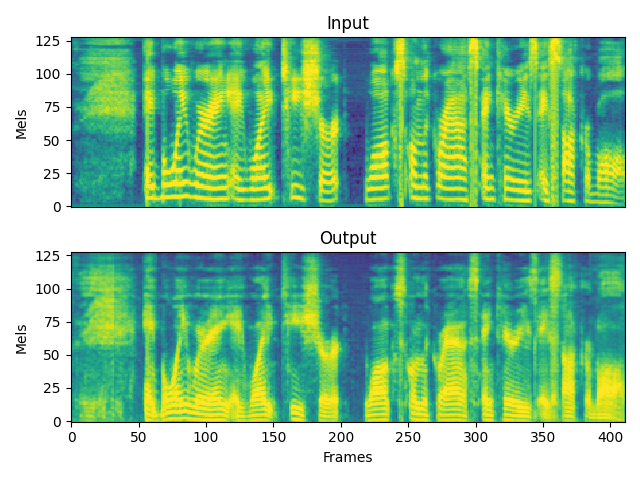}
    \caption{A target visualisation from the many-to-many experiment of male 1, with female 1 as the source input.}
    \label{fig:male1}
\end{figure}

\begin{figure}[h!]
    \centering
    \includegraphics[width=0.915\columnwidth]{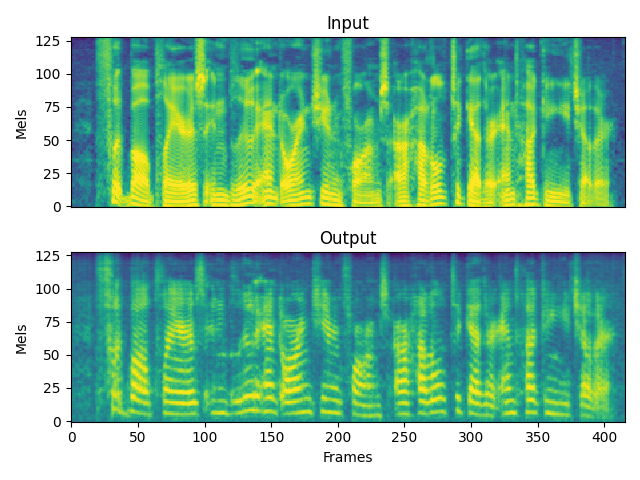}
    \caption{A target visualisation from the many-to-many experiment of female 2, with male 2 as the source input.}
    \label{fig:female2}
\end{figure}

\begin{figure}[h!]
    \centering
    \includegraphics[width=0.915\columnwidth]{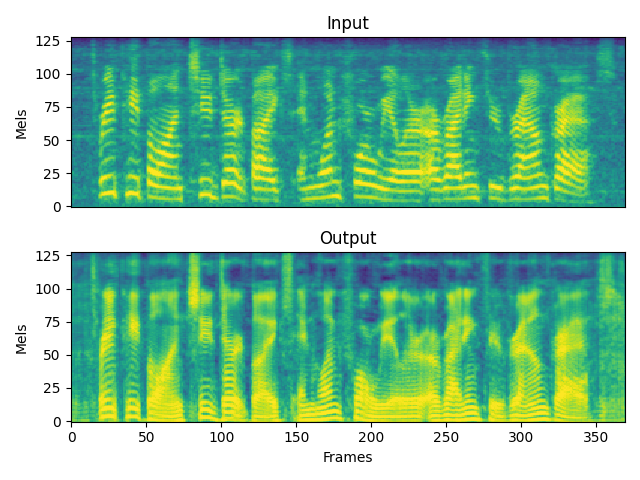}
    \caption{A target visualisation from the many-to-many experiment of male 2, with female 2 as the source input.}
    \label{fig:male2}
\end{figure}

\begin{figure}[h!]
    \centering
    \includegraphics[width=0.915\columnwidth]{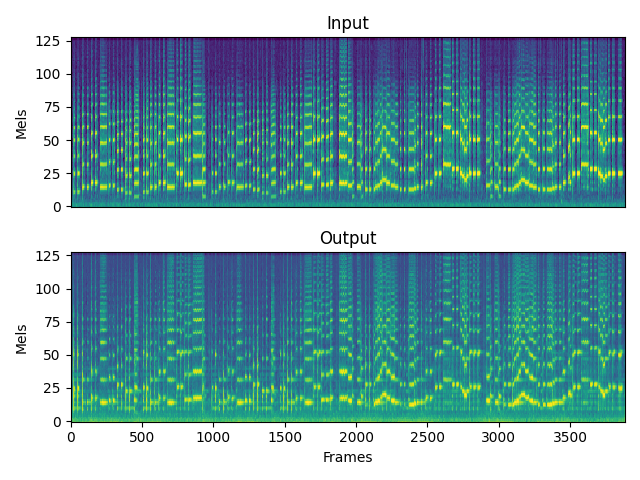}
    \caption{A target visualisation from the many-to-many experiment of violin, with trumpet as the source input.}
    \label{fig:violin}
\end{figure}

\begin{figure}[h!]
    \centering
    \includegraphics[width=0.915\columnwidth]{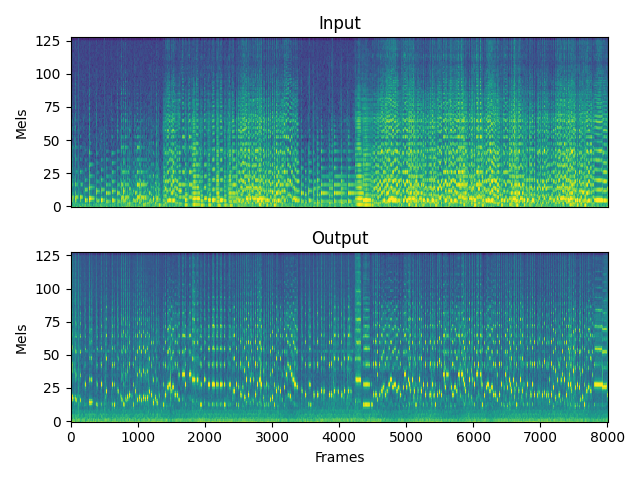}
    \caption{A target visualisation from the many-to-many experiment of flute, with cello as the source input.}
    \label{fig:flute}
\end{figure}

\begin{figure}[h!]
    \centering
    \includegraphics[width=0.915\columnwidth]{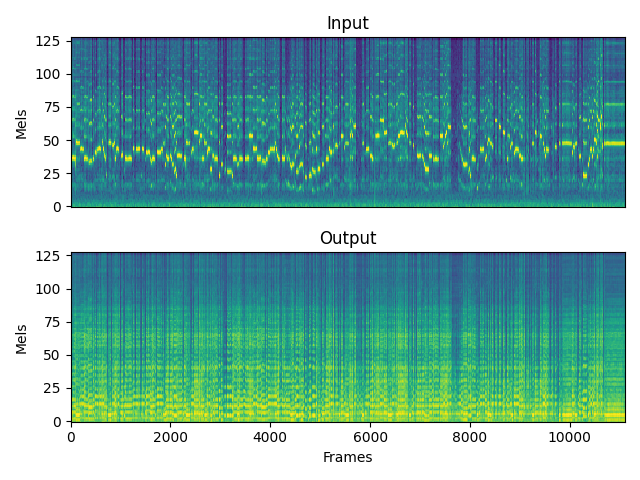}
    \caption{A target visualisation from the many-to-many experiment of cello, with flute as the source input.}
    \label{fig:cello}
\end{figure}

\subsection{WaveNet Vocoding Visualisations}
\label{sec:wavenet_vis}

\par
\begin{figure}[h!]
    \centering
    \includegraphics[width=\columnwidth]{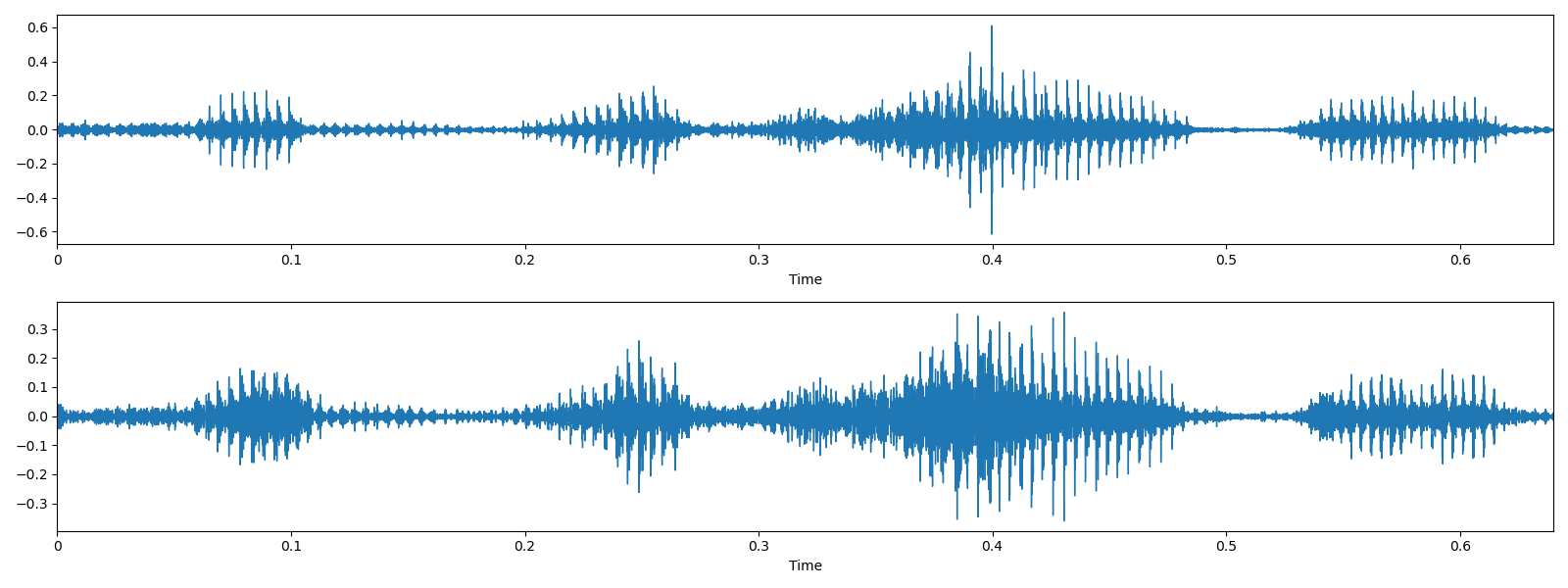}
    \caption{A vocoded reconstruction (bottom) of a real utterance (top) after training on speakers for 450,000 steps.}
    \label{fig:wavenet_flickr}
\end{figure}
\begin{figure}[h!]
    \centering
    \includegraphics[width=\columnwidth]{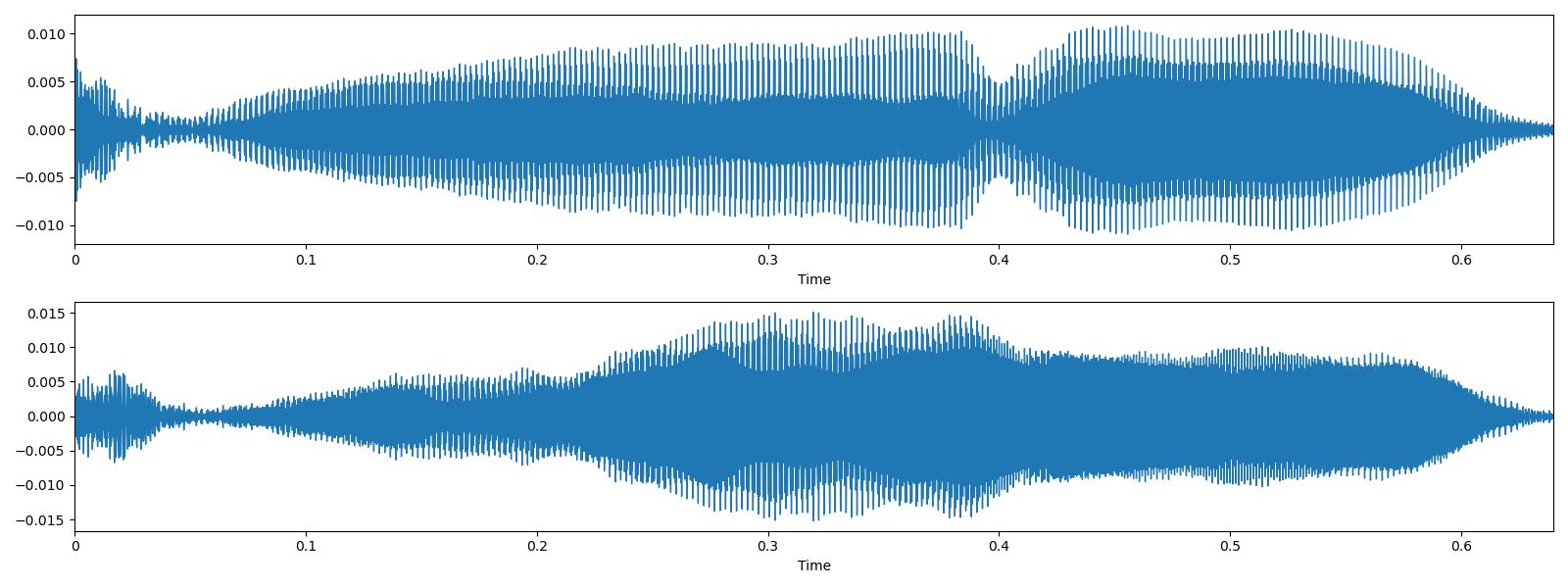}
    \caption{A vocoded reconstruction (bottom) of a real musical recording (top) after training on instruments for 450,000 steps.}
    \label{fig:wavenet_urmp}
\end{figure}

As presented in Figures \ref{fig:wavenet_flickr}-\ref{fig:wavenet_urmp}, after 450,000 steps two WaveNet vocoders are able to sufficiently reconstruct audio signals from mel-spectrograms for speakers and instruments, respectively. It is worth highlighting that the WaveNet vocoder for instruments reconstructs to less of an accurate degree for attacks and decays than the WaveNet vocoder for speakers as noticable at 0.05s and 0.4s of Figure \ref{fig:wavenet_urmp}. If this is due to the expressive complexity of the instruments, more steps or data per instrument would be appropriate for training in future experiments. For training with more steps, a faster time-frequency vocoder could also be considered.

\subsection{Hardware Specifications}

\par
The three most expensive parts of the project from least to most expensive were; the VAE-GAN, the WaveNet vocoder, and the FAD evaluation. In order to execute this project it is recommended to work with NVIDIA GPUs with VRAM as enlisted in Table \ref{tab:gpu}.

\vspace{0.4cm}

\begin{table}[h!]
\centering
\caption{GPU Specifications}
\label{tab:gpu}
\begin{tabular}{|c|c|}
\hline
\textbf{Stage} & \textbf{Recommended VRAM (GB)} \\ \hline
VAE-GAN        & 2                              \\ \hline
WaveNet        & 8                              \\ \hline
FAD            & 24                             \\ \hline
\end{tabular}
\end{table}

\end{document}